\documentclass[conference]{IEEEtran}
\usepackage{graphicx}
\usepackage{mathtext}
\usepackage{amssymb}

\begin{document}

\title{Generating Function For Network Delay}

\author{\IEEEauthorblockN{A.M. Sukhov, N.Yu. Kuznetsova, A.K. Pervitsky and A.A. Galtsev}
\IEEEauthorblockA{Samara State Aerospace University, Moskovskoe sh., 34, Samara, 443086, Russia\\
\em e-mails: amskh@yandex.ru, meneger$\_$job@mail.ru, pervitskiy.alex@mail.ru,  galaleksey@gmail.com}}

\maketitle

\begin{abstract}
In this paper correspondence between experimental data for packet delay and two theoretical types of distribution is investigated. Statistical tests have shown that only exponential distribution can be used for the description of packet delays in global network. Precision experimental data to within microseconds are gathered by means of the RIPE Test Box. Statistical verification of hypothesis has shown that distribution parameters remain constants during 500 second intervals at least. In paper cumulative distribution function and generating function for packet delay in network are in an explicit form written down, the algorithm of search of parameters of distribution is resulted.
\end{abstract}



\section{Introduction}
\label{intr}

The special area of the control theory, named networked control systems in which transfers as environment of operating signals were used computer networks, has arisen in the late nineties of the XX-th century~\cite{zbp}. Originally, as the network environment of control systems local networks~\cite{Georges05} which differ high-speed data transfer and in the minimum percent of packet loss were used.

The networked control systems in which as the handle environment the global network Internet is used, is extremely complicated because of random character of distribution of packet delay and their big absolute values~\cite{Hespanha07,tch,Zampieri08}. However till now, results of the advanced network researches are not used in the control theory and algorithms on their basis are not created. The present project assumes introduction of new network decisions in networked control systems.

Except classical problems of the control theory, there is requirement of management and monitoring of network processes. For example, for video transfer over networks TCP/IP networks the important parameter is the available bandwidth between two hosts~\cite{drm}. For problems of routing it is necessary to know throughput between routers. Special network emulators are applied to modeling of the majority of network processes, for example, INET/OMNET ++~\cite{var}.

Now numerous programs which imitate transmission of packets through TCP/IP protocol network are created. In a basis of operation of all emulators position that the type of delay distribution is unknown is occupied. The purpose of our research consists in that not only to define delay type, but also to find generating function for the traffic emulators.

For the decision of problems of the networked control systems on the basis of stack TCP/IP it is more convenient to use the process approach to the control theory, based on idea of existence of some universal functions of control. The purpose of our research is the finding of this function for network components. In the modern theory of computer networks there were many utilities working with a network delay, there is a progress in studying and modeling of transmitting of packets. Our problem consists in trying to describe process of a network delay of management packets and to show ways of practical calculation of all parameters entering into corresponding distribution functions~\cite{Fridman04}.

By transmission of control signals through TCP/IP network, the separate packets of the controlling data flow transferring the information are supplied non-uniformly, and the part of packets in general is lost by transmission on a network and does not reach a target. For rise of efficiency of control algorithms it is necessary to reduce to a minimum of packets delay and their variation, and also percent of packet loss. Similar algorithms are used for transmission voice and video streams, in grid systems, at control of robust systems, in network computer games, etc.

At first it would be desirable to result the brightest research on a distribution type for network delay. To understand, about what there is a speech in described papers, will give definitions of notations used in them:

\begin{itemize}
	\item 
	Round-trip time ({\em RTT\/}) time is the time required for a packet to travel from the testing host to a remote computer that receives the packet and retransmits it back to the source.
	\item
	The One-Way Delay ({\em OWD\/}) value is calculated between two synchronized points A and B of an IP network, and it is the time in seconds that a packet spends in travelling across the IP network from A to B.
\end{itemize}

In particular, Elteto and Molnar~\cite{Elteto99} have spent measurements of  round-trip delay  in the Ericsson Corporate Network, complex analysis of the received data has allowed to build the supposition about distribution type for network delay. The main finding of their research is that the round-trip delay can be well approximated by a truncated normal distribution.

Konstantina Papagiannaki {\em et al\/}~\cite{pmft} in the research have measured and have analyzed packet delay between two adjacent routers in the core network. On the basis of the received measurements, they have made the supposition about the factors influencing occurrence of delay, and very big delays which cannot be explained in the way of batch processing in routers on algorithm FIFO have been noticed.

Recently, fulfilling a series of operations on measurement of an available bandwidth~\cite{sss}, we have installed that for a type definition of delay distribution we should research only a variable part of delay while its most part remains constant. This fact also has served as a starting point of our operation.

\section{Premises for model}
\label{sb}

In 1999 Downey \cite{dow} for the first time has detected linear dependence of the minimum possible round trip time on the size of transferred packets.
In 2004 precise experiments by Choi et al \cite{chm}, Hohn et al~\cite{hvp} proved that the minimum fixed delay component $D^{fixed}(W)$ for a packet of size $W$ is a {\it linear} (or precisely, an {\it affine}) function of its size, 
\begin{equation}
  D^{fixed}(W)=W\sum_{i=1}^h 1/C_i + \sum_{i=1}^h \delta_i
  \label{eq2}
\end{equation}
where $C_i$ is each link of capacity of $h$ hops and $\delta_i$ is propagation delay. To validate this assumption, they check the minimum delay of packets of the same size for three path, and plot the minimum delay against the packet size. 

Let $D(W)$ represents the one way delay (point-to-point delay) of a packet. Here we refer to it as the minimum path transit time for the given packet size $W$, denoted by $D^{fixed}(W)=\min D(W)$. With the fixed delay component $D^{fixed}(W)$ identified, we can now subtract it from the point-to-point delay of each packet to study the variable delay component $d^{var}$. The variable delay component of the packet, $d^{var}$, is given by 
\begin{equation}
	D(W)=D^{fixed}(W)+ d^{var}
	\label{dvar}
\end{equation}
Computed minimal delay $D^{fixed}(W)$ is
\begin{equation}
	D^{fixed}(W)=D_{min}+W/C,
	\label{C-for}
\end{equation}
where $C=(W_2-W_1)/(D_2-D_1)$ is end-to-end available bandwidth and it is searched, comparing average time of packet delay of the different sizes $W_2$ and $W_1$~\cite{sss}. Here
\begin{equation}
	D_{min}=\lim_{W\rightarrow 0}D^{fixed}(W)
	\label{Dmin}
\end{equation}
The value $D_{min}$ is related to the distance between the sites (i.e. propagation delay) and per-packet router processing time at each hop along the path between the sites \cite{cml,cc}. This value represents as the minimum delay $D_{min}$ for which the very small package can be transmitted on a network from one point in another. 

The minimal delay~\cite{sss} of datagram transmission $D_{min}$ may be calculated as
\begin{equation}
  D_{min}=\frac{W_2D^{fixed}(W_1) - W_1D^{fixed}(W_2)}{W_2-W_1}
\end{equation}	
This value as well as the methods of its measurement has a important significance in applied tasks of control theory~\cite{zbp}. The second significant question of networking control theory is the distribution type for variable delay component $d^{var}$ which should be studied in rest of paper. To know the expression for this parameter we may easy calculate the duration of buffer for streaming application on receiving side.

\section{Experimental search}
\label{s2}

In order to determine distribution type for a variable delay component $d^{var}$ we should run considerable quantity of measurements between various hosts in the Internet. The basic problem of experimental testing is the precise of delay measurements that is necessary for accurate result. Similar measurements demand, at least, microsecond precision for delay measurements; we are reaching such accuracy with help of RIPE Test Box mechanism~\cite{ggk,ripe}. In order to prepare the experiments three Test Boxes have been installed in Moscow, Samara and Rostov on Don during 2006-2008 years in framework of RFBR grant 06-07-89074. Each RIPE Test Box represents a server under management of an FreeBSD operating system with the GPS receiver connected to it.

Characteristic times of investigated processes (a packet delay, jitter) have the order from 10 $ms$ to 1 $sec$, therefore is enough to use system hours of a RIPE Test Box for their reliable measurement. At the first stage experiment between tt01.ripe.net (RIPE NCC at AMS-IX, Amsterdam), tt143.ripe.net (Samara, SSAU), tt17.ripe.net (Bologna) and tt74.ripe.net (Melbourne) have been made which include precision measurement of packet delay with accuracy 2-12 $\mu s$. Testing results are available via telnet to corresponding RIPE Test Box on port 9142. It is important to come and write down simultaneously the data on both ends of the investigated connection. 

On the basis of the received data set it is easy to construct a cumulative distribution function for network delay $D$:
\begin{equation}
	F(D)=P(x\leq D)
	\label{del}
\end{equation}
 
For initial comparison truncated normal and exponential distributions have been chosen, expressions for which are written down.

For truncated normal distribution it is possible to select following approximation:
\begin{equation}
F(D)=\left\{ \begin{IEEEeqnarraybox}[][c]{rl}
0, & D< D_{min} ;\\
\frac{\sqrt{2/\pi}}{\sigma}\int\limits_{D_{min}}^D \exp \left\{-\frac{(x-D_{min})^2}{2\sigma^2}\right\}dx, & D \geq D_{min}
\end{IEEEeqnarraybox} \right.
\label{Dnor}
\end{equation}
where 
\begin{equation}
\sigma=D_{av}-D_{min}
\label{sigma}
\end{equation}
is the difference between average network delay $D_{av}(W)=\mathbb{E}[D(W)]$ and  minimum delay $D_{min}(W)$.

It should be noted that all statistical data has been gathered by us for the fixed size of a packets $W$. By default for RIPE Test Box it equals to 100 bytes. In Section~\ref{s6} we update a cumulative distribution function $F(D,W)$ taking into account the packets size $W$.

The alternative type of allocation which will be checked on correspondence is an exponential distribution, expression for which is written below. 
\begin{equation}
F(D)=\left\{ \begin{IEEEeqnarraybox}[][c]{rl}
0, & D< D_{min} \\
1-\exp\left\{-\lambda (D- D_{min})\right\},& D\geq D_{min}
\end{IEEEeqnarraybox} 
\right.
\label{Dexp}
\end{equation}
where 
\begin{equation}
\lambda=1/(D_{av}-D_{min}) 
\label{lambda}
\end{equation}
is reciprocal to the difference between average network delay $D_{av}(W)=\mathbb{E}[D(W)]$ and  minimum delay $D_{min}(W)$.

For initial check of conformity to distribution type two methods will be used: calculation of Pearson correlation coefficient and a graphic method. We will designate as $K_{nor}$ correlation coefficient between experimental and normal distributions then $K_{exp}$ is correlation coefficient between experimental and exponential distributions. 

The obtained data is shown in Table~\ref{prD}, where the column {\bf host} corresponds to a direction between two RIPE Test Boxes, and the column $W$ specifies in the size of a testing packet.

\begin{table}
	\centering
		\begin{tabular}{|l|l|c|c|c|} \hline
N & host & $W$ (bytes) & $K_{nor}$ & $K_{exp}$ \\ \hline
1 & bolonia &&&\\
  & tt01$\Rightarrow$tt17 & 100 &0.76 & 0.97 \\ \hline
2 & samara    &&& \\
  & tt01$\Rightarrow$tt143 & 100 & 0.87 & 0.98 \\ \hline
3 & samara   &&& \\
  & tt01$\Rightarrow$tt143 & 1024 & 0.99 & 0.99 \\ \hline
4 & melburn  & && \\
  & tt01$\Rightarrow$tt74 & 100 & 0.66 & 0.97 \\ \hline  
	\end{tabular}
	\caption{Precise measurements}
	\label{prD}
\end{table}

\begin{figure}
\centering
\includegraphics[height=4cm]{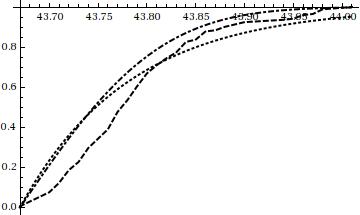}
\caption{Experimental (dash), normal (dash-dot) and exponential (dot) CDFs, precise testing. Direction: tt01$\Rightarrow$tt143, $W=$100 bytes}
\label{f1a}
\end{figure}

\begin{figure}
\centering
\includegraphics[height=4cm]{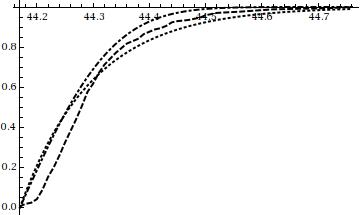}
\caption{Experimental (dash), normal (dash-dot) and exponential (dot) CDFs, precise testing. Direction: tt01$\Rightarrow$tt143, $W=$1024 bytes}
\label{f1b}
\end{figure}

Except correlation coefficients it is possible to compare graphics representation of cumulative distribution functions (CDF), showing all three functions on one plot. On the uniform graphics (see Figures~\ref{f1a}, \ref{f1b}) dash line selects an experimental curve, dot-dash curve corresponds to normal allocation. In dot style the exponential distribution is painted.

The plots constructed in Figures 1 and 2, represent dependence of CDF on delay of a packet on a site from Amsterdam to Samara (tt01$\Rightarrow$tt143). The first plot describes testing of a network by packages in the 100 bytes size, the second plot corresponds to packages in the size of 1024 bytes. Time on axis $Õ$ is measured in milliseconds.

All experiments resulted above testify that the preferable type of distribution describing packet delay in a global network is an exponential distribution. Thus, as have shown our researches, the random variable of packet delay between two network points is arranged on an exponential low with the parameter calculated from experimental values under the Equation~(\ref{lambda}).

However, not each investigator who is engaged in the control theory has the equipment for the precision measurements, similarly RIPE Test Boxes. Therefore in this part it would be desirable to present technique which uses the data of well-known utilities and doesn't demand the expensive equipment.

For testing it is possible to use the utility {\em ping} as it is the most widespread resource for verification of  connection quality in TCP/IP networks. Let's mark only that this utility measures round-trip time, instead of one way delay.

The data received with help of {\em ping} has the millisecond precision that is exact enough to judge delay distribution. The utility {\em ping} allows testing connections between points AIST - New Zealand (tt47.ripe.net), Volgatelekom - Australia (tt74.ripe.net) and SSAU-Melbourn (tt74.ripe.net). As remote hosts were used servers of RIPE measurement system, AIST, Volgatelecom (VT), Infolada and SSAU is local Internet Service Providers from Samara region, Russia. Processing the obtained data on the above described algorithm, we have received the results presented in the Table~\ref{pingD}.

\begin{table}
	\centering
		\begin{tabular}{|l|l|c|c|c|} \hline
N & host & $W$ (bytes) & $K_{nor}$ & $K_{exp}$ \\ \hline
1 & AIST$\Rightarrow$ &&&\\
  & New Zeland & 32 & 0.94 & 0.95 \\ \hline
2 & Volgatelecom$\Rightarrow$    &&& \\
  & Australia & 32 & 0.96 & 0.98 \\ \hline
3 & SSAU$\Rightarrow$  &&& \\
  & Melburn & 64 & 0.66 & 0.97 \\ \hline
4 & Infolada$\Rightarrow$  & && \\
  & Athens & 32 & 0.98 & 0.98 \\ \hline  
	\end{tabular}
	\caption{{\em ping} measurements}
	\label{pingD}
\end{table}

\begin{figure}
\centering
\includegraphics[height=4cm]{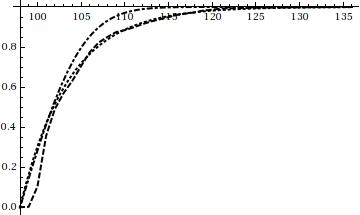}
\caption{Experimental (dash), normal (dash-dot) and exponential (dot) CDFs, precise testing. Direction: Samara$\Rightarrow$Holland, $W=$32 bytes}
\label{f3}
\end{figure}

\begin{figure}
\centering
\includegraphics[height=4cm]{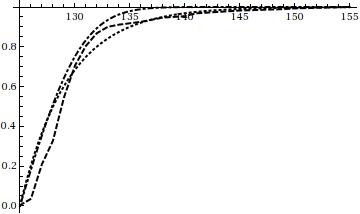}
\caption{Experimental (dash), normal (dash-dot) and exponential (dot) CDFs, precise testing. Direction: Infolada$\Rightarrow$Athens, $W=$32 bytes}
\label{f7}
\end{figure}

\begin{figure}
\centering
\includegraphics[height=4cm]{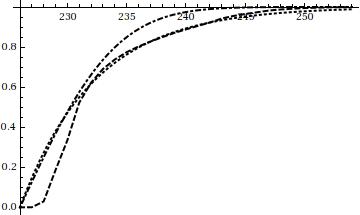}
\caption{Experimental (dash), normal (dash-dot) and exponential (dot) CDFs, precise testing. Direction: SSAU$\Rightarrow$Australia, $W=$1064 bytes}
\label{f11}
\end{figure}

The evident illustration is resulted in definition of distribution type in Figure~\ref{f3}-\ref{f11}.

It should be noted that the utility {\em ping\/} allows finding automatically values of variables $D_{av}$ and $D_{min}$ (see Eqns.~(\ref{Dnor}) and (\ref{Dexp})) which completely define the distribution form, both normal and exponential types. It is enough to give sequence from 20 packets to obtain the given values with split-hair accuracy, sufficient for the description of processes of the control theory.

\section{Statistical hypothesis testing}
\label{s4}

The verification executed in the previous section about conformity of distribution type has preliminary character and isn't strict. In this section for check of distribution type Pearson's chi-square test will be used.

For processing some data sets from 2000 to 2500 values the delays collected with interval in 2 seconds have been used. This data divides into intervals in 50, 100, 250, 500, 1000 and 2000 values and was tested on Pearson. Testing results could be found in Tables~\ref{t3},~\ref{t4}.

At construction of these Tables following designations were used:
\begin{itemize}
	\item Dimension of observations $N$ (number of measurements)
	\item $n$ is the number of cells. All observations $N$ are divided among $n$ cells according Sturges' rule $n=(1+3.22\lg N)+1$
	\item $t$ is the value of the test-statistic
	\item $\chi^2_{0.95,n-1}$ is the theoretical value of threshold of hypothesis acceptance
\end{itemize}
If $t<\chi^2_{0.95,n-1}$  then hypothesis about corresponding type of distribution is accepted, differently hypothesis is rejected.

In the beginning we will check up on conformity to exponential distribution the data, reception by means of RIPE Test Box, see Table~\ref{t3}.

\begin{table*}
	\centering
		\begin{tabular}{|l|c|c|c|c|c|c|c|} \hline
$N$ & 50 &	100 &	200 &	250 & 500 &	1000 & 2000\\ \hline
$n$ &	14 & 17 &	19	& 20 &	22 &	24 &	27\\
$\chi^2_{0.95,n-1}$ & 22.36 &	26.30 &	28.87 &	30.14 &	32.67 &	35.17 &	38.89\\
$t$ &	21.29 &	25.35 &	22.77 &	23.10 &	134.31 &	547.16 &	978.98\\
Hypothesis acceptance & Yes & Yes & Yes & Yes & No & No & No \\ \hline  
	\end{tabular}
	\caption{Verification of exponential distribution, Samara-Amsterdam, (tt143$\Rightarrow$tt01), packet size 100 bytes}
	\label{t3}
\end{table*}

From Table~\ref{t3} follows that within 500 second intervals (250 measurements) the packet delay is distributed on the exponential law.

The data received with RIPE Test Box, was checked also on conformity to the truncated normal distribution, see Table~\ref{t4}.

\begin{table*}
	\centering
		\begin{tabular}{|l|c|c|c|c|c|c|c|} \hline
$N$ & 50 &	100 &	200 &	250 & 500 &	1000 & 2000\\ \hline
$n$ &	14 & 17 &	19	& 20 &	22 &	24 &	27\\
$\chi^2_{0.95,n-1}$ & 22.36 &	26.30 &	28.87 &	30.14 &	32.67 &	35.17 &	38.89\\
$t$ & 43.32 &	217.46 &	2906.47	& 6077.07 & $\infty$ & $\infty$  & $\infty$ \\
Hypothesis acceptance & No & No & No & No & No & No & No \\ \hline  
	\end{tabular}
	\caption{Verification of truncated normal distribution, Samara-Amsterdam, (tt143$\Rightarrow$tt01), packet size 1024 bytes}
	\label{t4}
\end{table*}

Pearson's chi-square test allows rejecting hypothesis about the truncated normal type of distribution for the description of process of packet delay.

\section{Distribution Type for Delay and Generating Function for Traffic Emulator}
\label{s6}

In real the Internet processes the size of transferred packages can vary, therefore the cumulative distribution function should be updated. For each size of a packet $W$ there is the minimum time $D^{fixed}(W)$ defined by the Eq.~(\ref{C-for}).

Then, the final cumulative distribution function $F(D,W)$ is
\begin{equation}
F(D, W)=\left\{
\begin{IEEEeqnarraybox}[][c]{rl}
0, D < D_{min}+W/C &\\
1-\exp\left\{-\lambda (D- D_{min}-W/C)\right\}, &\\  D\geq D_{min}+W/C &
\end{IEEEeqnarraybox}
\right.
\label{DexpF}
\end{equation}
where 
\begin{equation}
\lambda=1/(D_{av}-D_{min}) 
\label{lambda1}
\end{equation}
is reciprocal to the difference between average network delay $D_{av}(W)=\mathbb{E}[D(W)]$ and  minimum delay $D_{min}(W)$ and $C$ is end-to-end capacity.

It is important to notice that the data on the delay, received with use of packages of various length, comprises the additional variable component caused network jitter $j$ (delay variation)~\cite{drm}. Therefore, the best the controlling algorithm will form packages of the identical size. If we use the utility {\em ping \/} for delay definition in it there is a special key for resizing of a testing package ($-l$ in Windows, $-s$ in Linux).

Our model has one more wide application except tasks of the control theory. Results of our operation can be applied at writing of emulators of the traffic in a global network~\cite{var}. Till now it was considered that the type of delay distribution is unknown, and traffic emulators used own generating functions for delay generation. On the basis of the type of delay found us (see Eqn.~(\ref{DexpF})) it is possible to write generating function
\begin{equation}
D=D_{min}+W/C -(1/\lambda)\ln(1-F(D,W))
\label{gf}
\end{equation}

In this equation content distribution function (CDF) $F(D,W)$ can be set the generator of random numbers. The received numbers will give values of delay for network packet of the different size. We will notice once again that in real networks these values can be calculated according to the specilised utility.

\section{Conclusion}
\label{s7}

In the present work for the description of process of the packet delay in a global networks it has been chosen exponential distribution. In comparison with truncated normal distribution it has shown the best correlation with experimental results. Unlike the truncated normal distribution it passes check by statistical methods. 

Experimental data were gathered by means of the precision RIPE measuring system to within microseconds, and also by means of the standard utility {\em ping\/}. This utility measures round-trip time to within milliseconds. During small periods about several minutes when it is possible to consider conditions of transmission on a TCP/IP network invariable, such approach gives correlations from above 0.99. At change of network conditions the elementary {\em ping\/} testing by a series from 20 packets will allow to change exponential distribution parameters instantly.

We write in an explicit form of cumulative distribution functions for normal and exponential distribution of delay. Generating function for packet delay has been found in real networks which can be used in emulators of the traffic of a global network.

In summary we would like to thank Leonid Fridman, the Professor from University of Mexico for fruitful dialogues in which course the idea of this article has taken shape. Also it would be desirable to thank all collective of technical service RIPE ncc and especially Ruben van Staveren and Roman Kalyakin for constant assistance in comprehension of subtleties of a measuring infrastructure. We also would like to express the gratitude to the Wolfram Research corporation, which the first has marked our preprint and has given us licenses for the right of use of 
{\it Mathematica}

\end{document}